\documentclass[11pt,twoside]{article}
\usepackage{asp2004}
\usepackage{epsf}
\usepackage{psfig}
\begin{document}
\title{The Photometric Calibration of the Dark Energy Survey}
\author{Douglas L.\ Tucker, James T.\ Annis, Huan Lin, Stephen Kent, Chris Stoughton, John Peoples, Sahar Allam}
\affil{Fermilab, MS~127, P.O. Box 500, Batavia, IL 60510-0500, USA}
\author{Joseph J.\ Mohr, Wayne A.\ Barkhouse, Choong Ngeow}
\affil{Dept.\ of Astronomy, University of Illinois, Urbana, IL 61801, USA}
\author{Tanweer Alam, Cristina Beldica, Dora Cai, Greg Daues, Ray Plante}
\affil{National Center for Supercomputing Applications, Urbana, IL 61801, USA}
\author{Chris Miller, Chris Smith}
\affil{Cerro Tololo Inter-American Observatory/NOAO, 950 N.\ Cherry Ave., AZ 85719 USA}
\author{Nicholas B.\ Suntzeff}
\affil{Physics Department, Texas A\&M University, College Station, TX 77843-4242 USA)}
\author{For the Dark Energy Survey Collaboration}

\begin{abstract} 
The Dark Energy Survey (DES) is a 5000 sq deg $griz$ imaging survey to
be conducted using a proposed 3 sq deg ($2.2\deg$-diameter) wide-field
mosaic camera on the CTIO Blanco 4m  telescope. The primary scientific
goal of  the DES is to  constrain dark  energy cosmological parameters
via four complementary methods: galaxy cluster counting, weak lensing,
galaxy   angular correlations,  and Type  Ia  supernovae, supported by
precision photometric  redshifts.  Here  we  present the   photometric
calibration plans for the   DES, including  a discussion  of  standard
stars and field-to-field calibrations.
\end{abstract}

\section{Introduction}

The goal of the Dark  Energy Survey (DES) is  to perform a 5000 sq deg
$griz$  imaging  survey of  the Southern   Galactic   Cap in order  to
constrain the  Dark   Energy equation-of-state    parameter $w  \equiv
P/\rho$  to    $\sim$5\%  (statistical  errors)  in    each  of   four
complementary techniques and  to begin to  contrain  the derivative of
$w$ with respect  to redshift ($dw/dz$).    By meeting this  goal, DES
will  serve as a  stepping  stone to next-generation large-scale  dark
energy   projects   like  Large   Synoptic   Survey   Telescope (LSST;
\citealt{Sweeney06}),      the      Joint    Dark    Energy    Mission
(JDEM)\footnote{{\tt
http://spacescience.nasa.gov/admin/divisions/sz/SEUS0310/Hertz\_JDEM.pdf}},
and the Square Kilometer Array (SKA; \citealt{Beck05}).

The   survey will  utilize the Blanco    4m telescope at Cerro  Tololo
Interamerican Observatory (CTIO).  In  order  to achieve the  goal  of
surveying half of the Southern sky in a reasonable amount of time, the
DES collaboration will replace Blanco's prime  focus cage with a new 3
sq  deg ($2.2\deg$-diameter) optical  CCD  mosaic imager known as  the
Dark Energy Camera  (DECam).  The  DECam  is scheduled for design  and
construction during the 2005--2009 time period and, once commissioned,
will be available for community use as well as for the DES.

Observations for the DES will encompass a total of 525 nights (30\% of
the telescope time) over the course  of 5 years starting in 2009/2010.
For best viewing  of the Southern Galactic  Cap,  the DES observations
will be performed during the months of September through February.

\section{DES Science:  Four Probes of Dark Energy}

\begin{figure}    
\centering
\vspace{-.25cm}
\makebox[80mm]{\psfig{file=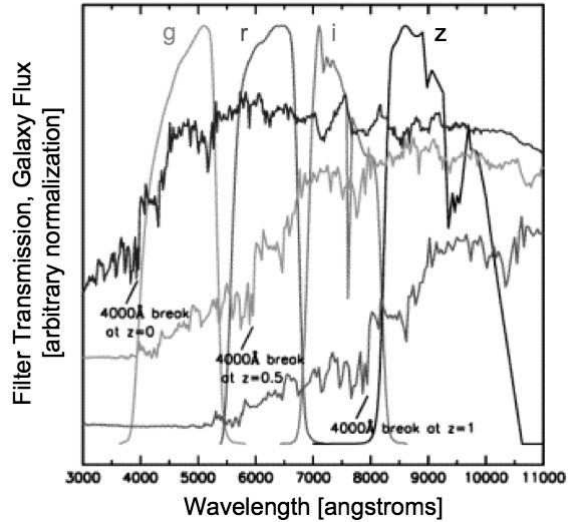,width=80mm,clip=,angle=0,silent=}}\vspace{-.25cm}
\caption{The response functions of  the DES $griz$ filters overplotted
on the spectral energy distributions of a typical elliptical galaxy at
$z=0$, $z=0.5$, and $z=1.0$.  Note that the passage of the 4000\AA
break through the different filters with increasing redshift and the
relative uniformity of the shapes of elliptical galaxy spectra permit
accurate photometric redshifts ($\sigma_z =$ 0.02--0.03 to be measured
for individual cluster galaxies. (Credit: Huan Lin)
\label{fig1}}
\end{figure}

The DES will measure the dark energy equation-of-state parameter $w$
to $\sim$5\% in each of four complementary probes:
\begin{description}
\item[Galaxy cluster counting:]  
	The DES will identify and  measure the redshifts and masses of
	20,000  galaxy   clusters  with  masses   $M >  2\times10^{14}
	M_{\sun}$ out to a redshift of $z=1.3$. The number of clusters
	as a function of redshift is a strong function of cosmological
	parameters, including $w$ \citep{Wang98,Haiman01}.
\item[Weak lensing:]  
	The DES will  identify and measure the  shapes of 300  million
	galaxies over  5000 sq  deg.   The  clustering  of mass  as  a
	function  of redshift  is also strongly  determined by various
	cosmological parameters, including $w$ \citep{Hu02,Huterer02}.
\item[Spatial clustering of galaxies:]  
	The DES will identify and measure the photometric redshifts of
	300 million galaxies  out to a  redshift of  $z=1$ and beyond.
	The spatial  clustering of galaxies  as a function of redshift
	is also a function of $w$ \citep{Cooray01}.
\item[Standard Candles:]  
	$\approx$10\% of   DES  observing time  will be  devoted  to a
	supernova survey  covering 40 sq deg  of sky.   It is expected
	that DES will discover 1900 Type Ia supernovae in the redshift
	range $z=0.25-0.75$ during the course of its supernova survey.
	Construction  of a Hubble Diagram using  Type Ia supernovae as
	standard  candles is   a   well-known means of   measuring the
	parameters of dark energy \citep{Riess98,Perlmutter99}.
\end{description}

Note that, since the DES is purely an imaging survey, good photometric
redshifts out   to a redshift  of  $z  \sim 1.3$ are  critical  to its
success (see   Fig $\ref{fig1}$).   To  obtain  sufficiently  accurate
photometric redshifts to achieve the DES goals, the all-sky photometry
must be accurate to 2\% or  better.  Such photometric accuracy is also
necessary to obtain good light curves for the supernova survey part of
the DES program.

\section{Basic Survey Parameters}

\begin{figure}    
\centering
\vspace{-.25cm}
\makebox[80mm]{\psfig{file=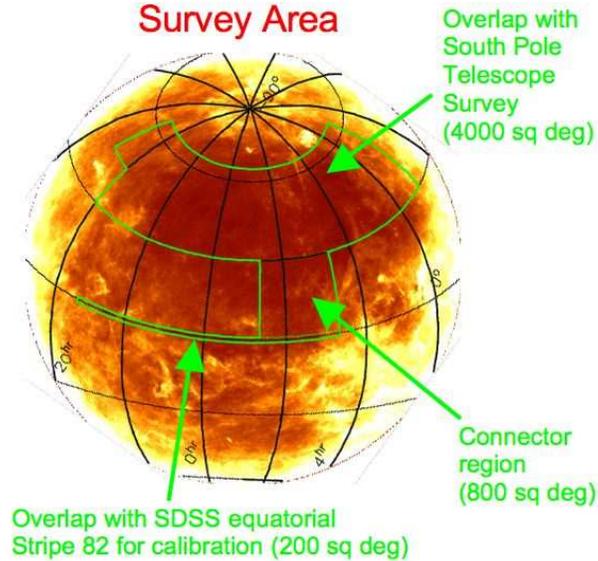,width=80mm,clip=,angle=0,silent=}}\vspace{-.25cm}
\caption{An equal area projection of the Southern Galactic Cap. 
A black body color scaling indicates interstellar extinction from the
Schlegel et al.(1998) dust maps (dark equals low extinction, bright
equals high extinction). The region delineated in green is the
proposed survey area for the DES.  (Credit:  Jim Annis)
\label{fig2}}
\end{figure}

The  DES  will not   only be  a wide survey    (5000 sq  deg;  see Fig
$\ref{fig2}$),   but  it  will also be   deep.    For galaxies,  it is
estimated that the   10$\sigma$ detection limit  will   be 24.6, 24.1,
24.3,  23.9 for $g$, $r$,  $i$,  and $z$, respectively.  Likewise, for
point sources, it is estimated that the 5$\sigma$ detection limit will
be 26.1, 25.6, 25.8, 25.4 for $g$, $r$, $i$, and $z$.  For comparison,
the Sloan Digital Sky Survey (SDSS)  reports a 95\% completeness limit
for point sources of $r=22.2$ \citep{Adelman-McCarthy06}.

The  observing strategy  is to   use 100  sec exposures,  and  observe
(typically) two  filters per telescope   pointing.  During dark  time,
these filters will be  $g$ and $r$; during  bright time, they will  be
$i$ and $z$, which are much less  affected by sky brightness caused by
moonlight.   Multiple tilings\footnote{A  tiling  is a single complete
coverage of the 5000   sq deg of the survey   area in one  filter} and
large offset overlaps will allow the DES  to achieve its desired depth
and will  be important  in optimizing photometric   calibrations.  The
current  idea for survey  strategy is to complete  two full tilings in
each filter each year, for a  five-year final tally  of 5 tilings each
in $g$ and $r$, 7 tilings in $i$, and 13 tilings in  $z$.  As noted in
the previous section,    the DES requirement for  all-sky  photometric
accuracy is 2\% (0.02 mag); an enhanced goal is 1\% (0.01 mag).

\section{The DES Instrument (DECam)}

\begin{figure}    
\centering
\vspace{-.25cm}
\makebox[60mm]{\psfig{file=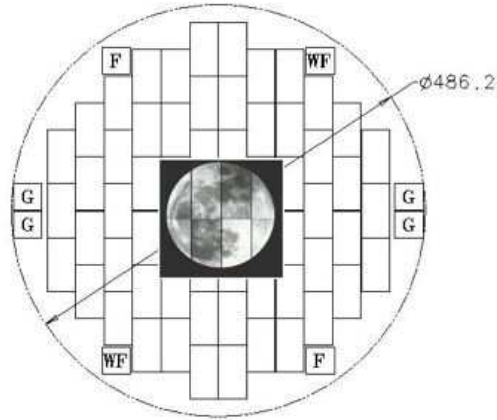,width=70mm,clip=,angle=0,silent=}}
\caption{The DECam focal plane.  To illustrate its field-of-view, an 
image of the full moon reproduced to scale is superimposed on the DECam
mosaic.  (Credit:  Brenna Flaugher)
\label{fig3}}
\end{figure}

The DECam focal plane (a.k.a., ``the Hex''; Fig.~\ref{fig3}) will be a
mosaic  containing 62  2k$\times$4k  imaging  CCDs, yielding a  camera
having a grand total of 520 million pixels.  At the prime focus of the
Blanco,  the   pixel scale   of   this  camera   will be  0.27  arcsec
pixel$^{-1}$.   The CCDs themselves   are a Lawrence Berkeley National
Laboratory (LBNL) design.  They  are  fully depleted and  250  microns
thick, yielding high quantum  efficiency in the $z$  band (QE $>$ 50\%
at 1 micron).  The  CCD electronics will permit  the full camera to be
read out in only 17 seconds.

\section{DES Calibrations Flow Diagram}

\begin{figure}    
\centering
\vspace{-.25cm}
\plotone{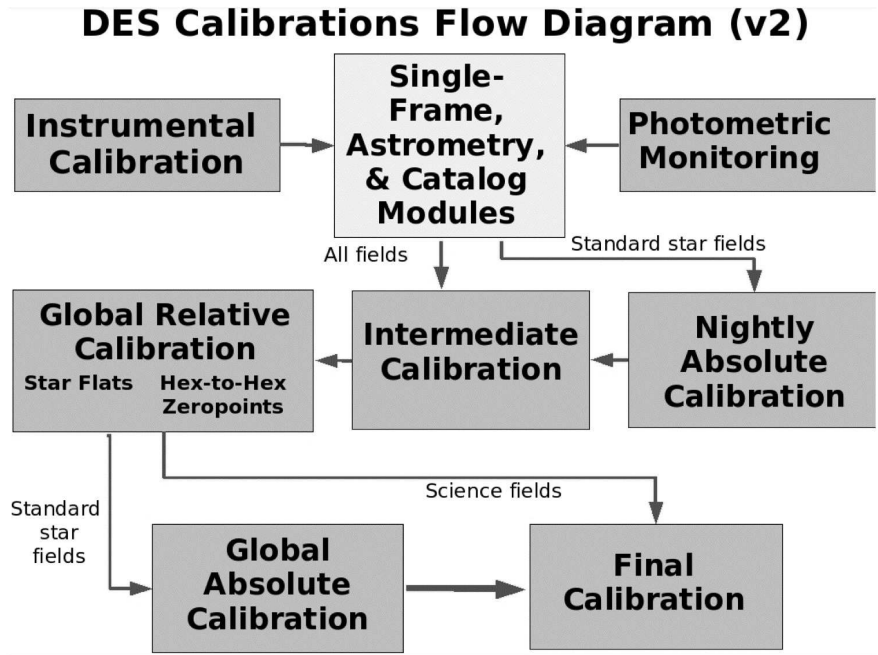}
\caption{Proposed plan for DES Calibrations.\label{fig4}}
\end{figure}

In Figure~\ref{fig4}  we   present a flowchart  which   represents our
current vision for the photometric calibration of  the DES.  The eight
boxes represented therein are as follows:

\begin{description}

\item[Instrumental Calibration:] 
This box contains those aspects  of calibration that are necessary for
image processing.  It  is here where raw bias  frames, dome flats, and
twilight flats are  processed and   combined  to create  master   bias
frames, dome flats and   twilight flats, respectively.  (Although  the
thickness of the LBNL CCDs may mean that $i$ and $z$ fringing will not
be a problem, it is also here where fringe frames would be created, if
necessary.)  This  box is also the domain  where scattered light maps,
shutter timing maps, CCD linearity curves, and chip-to-chip cross-talk
corrections would be be created.

\item[Photometric Monitoring:] 
This box  contains the apparatus  for independent real-time monitoring
of the photometric quality of   the sky.  Inputs could include  images
from  a  10 micron  all-sky  camera, images   from an  optical all-sky
camera,  or flux  measurements  from an  automated  differential image
motion monitor  (DIMM).  Outputs would include  diagnostic information
output to  a  webpage during   observations  (for real-time  observing
strategy decisions) and a photometricity flag added to the FITS header
of each image (for off-line  decisions  relevant to data  processing).
We describe Photometric Monitoring in more detail in Section 6 below.

\item[Single-Frame, Astrometry, \& Catalog Modules:]  
Here, the raw science and standard star frames are processed using the
master calibration frames,   scattered light and shutter  timing maps,
CCD linearity curves, and cross-talk  correction coefficients from the
Instrumental Calibration  box.  Output  is  a catalog  of objects from
each processed science  and standard star  frame with measured RA, DEC
positions   and  instrumental magnitudes   (as well  as other measured
properties  not  necessarily  of direct  importance   to calibration).
Information from Photometric  Monitoring concerning the sky conditions
at the time an image was taken will be carried along with the catalogs
generated from that  image.  This information  will aid in determining
whether or not data from a that  image is used  downstream of this box
for photometric calibration.  The Single-Frame, Astrometry, \& Catalog
Modules box represents the main image  processing and object detection
portion  of the DES data processing  pipeline; for more information on
the DES data processing pipeline, see \citet{Ngeow06} and
\citet{Mohr07}.

\item[Nightly Absolute Calibration:] 
Nightly  Absolute Calibration simply refers  to the process of fitting
the nightly standard star  observations  to a photometric equation  to
solve for (for instance)   the   nightly photometric zeropoints    and
atmospheric extinction  coefficients.  Input includes the instrumental
magnitudes  of   standard stars  observed   on  a given   night, their
airmasses, and  their   known standard magnitudes    and  colors.  The
photometricity  flag from Photometric Monitoring  can be  used to cull
out standard   star   observations   obtained under    non-photometric
conditions.  Output from  this box includes the photometric zeropoints
and   atmospheric  extinction coefficients  to  place the instrumental
magnitudes onto  the   AB  magnitude  scale \citep{Fukugita96}.     We
describe Nightly  Absolute  Calibration in  more detail  in  Section 7
below.

\item[Intermediate Calibration:] 
Intermediate   Calibration merely refers  to the  step of applying the
photometric   zeropoints and   extinction terms   measured  in Nightly
Absolute Calibration to  all the observations for  a given night.  The
result  is   calibrated magnitudes  for  all the  observations on that
night.     Since these   calibrated   magnitudes   may be  ``tweaked''
downstream  of   this box,   we refer to     this step as Intermediate
Calibration.

\item[Global Relative Calibration:] 
Global  Relative  Calibration  concerns measuring  two   tweaks to the
calibrated magnitudes from  the Intermediate Calibration step,  one of
which is a correction to any uncorrected vignetting or scattered light
effects in the photometry (Star Flat analysis)  and the other of which
is   the  calculation  of   relative  field-to-field  (``Hex-to-Hex'')
photometric   zeropoint   offsets.   We    describe   Global  Relative
Calibration in more detail in Section 9 below.

\item[Global Absolute Calibration:] 
Applying  the  zeropoint    adjustments   from the  Global    Relative
Calibration step might  offset the  observed standard star  magnitudes
from  their catalog  values.    The Global Absolute  Calibration  step
calculates the final overall zeropoints  needed --- one zeropoint  per
filter ---  to place  the  global relative  calibrations onto   the AB
magnitude scale.  We   describe  Global Absolute Calibration in   more
detail in Section 10 below.

\item[Final Calibration:] 
Here,  the zeropoint adjustments  from the Global Relative Calibration
and the Global  Absolute Calibration step are  applied to  the science
field observations.  The output   is calibrated AB magnitudes  for all
the science observations.

\end{description}

After the Final  Calibration step,  the  DES data should  be ready for
image and/or catalog co-addition to achieve  final images and catalogs
of the depth  required by the  DES science goals.  Barring problems in
bookkeeping --- and, in particular,  barring problems in keeping track
of photomeric zeropoint offsets --- there should be no need of further
calibration steps.   Thus, we do  not discuss any further steps beyond
Final Calibration in this paper.

Let us now discuss some of the above boxes in further detail, starting
with Photometric Monitoring.

\section{Photometric Monitoring}

Not all DES imaging will be done under  photometric conditions.  Those
images observed under photometric  conditions will be useful both  for
increasing the magnitude limit  of the final  co-added survey and  for
photometric calibration;  those  images observed under non-photometric
conditions will  only be useful for increasing  the magnitude limit of
the final survey.

Therefore,  photometric   monitoring    is    necessary.   Photometric
monitoring should  provide  real-time estimates of  sky conditions for
survey strategy.    In other  words, real-time  photometric monitoring
should be  able to answer the  question, ``Should the  next image be a
photometric  calibration (standard star) field,   a science target, or
something else?''

Photometric    monitoring  should also  provide  a    quality check or
photometricity flag that is usable offline during data processing.  In
this case, photometric monitoring should be  able to provide some sort
of measure  of the photometric quality of  an image already  taken and
thus be  able to answer the question,  ``This image was obtained under
such conditions;  is  it  good  enough to   be used   for  photometric
calibrations?''

\begin{figure}    
\centering
\vspace{-.25cm}
\plottwo{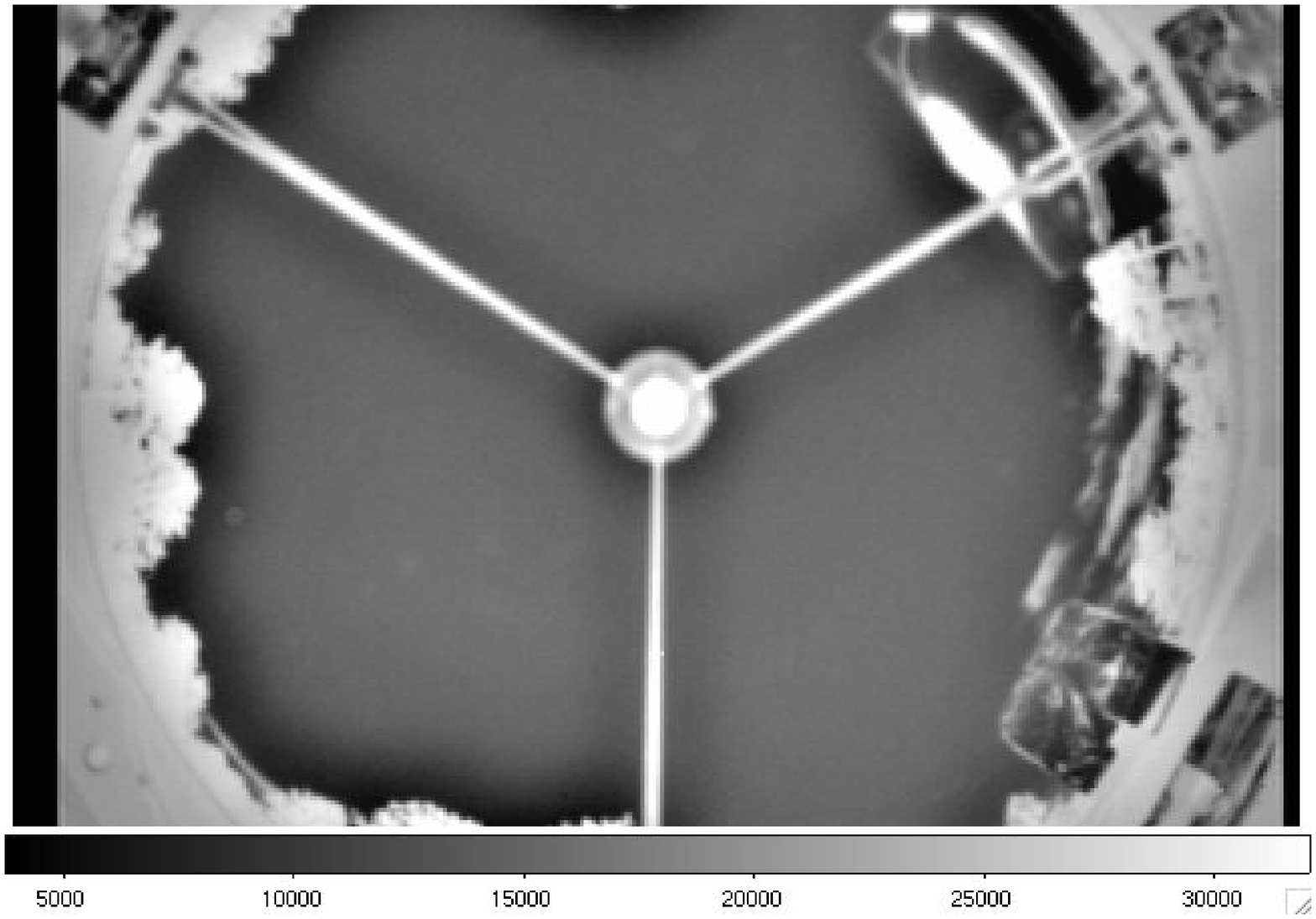}{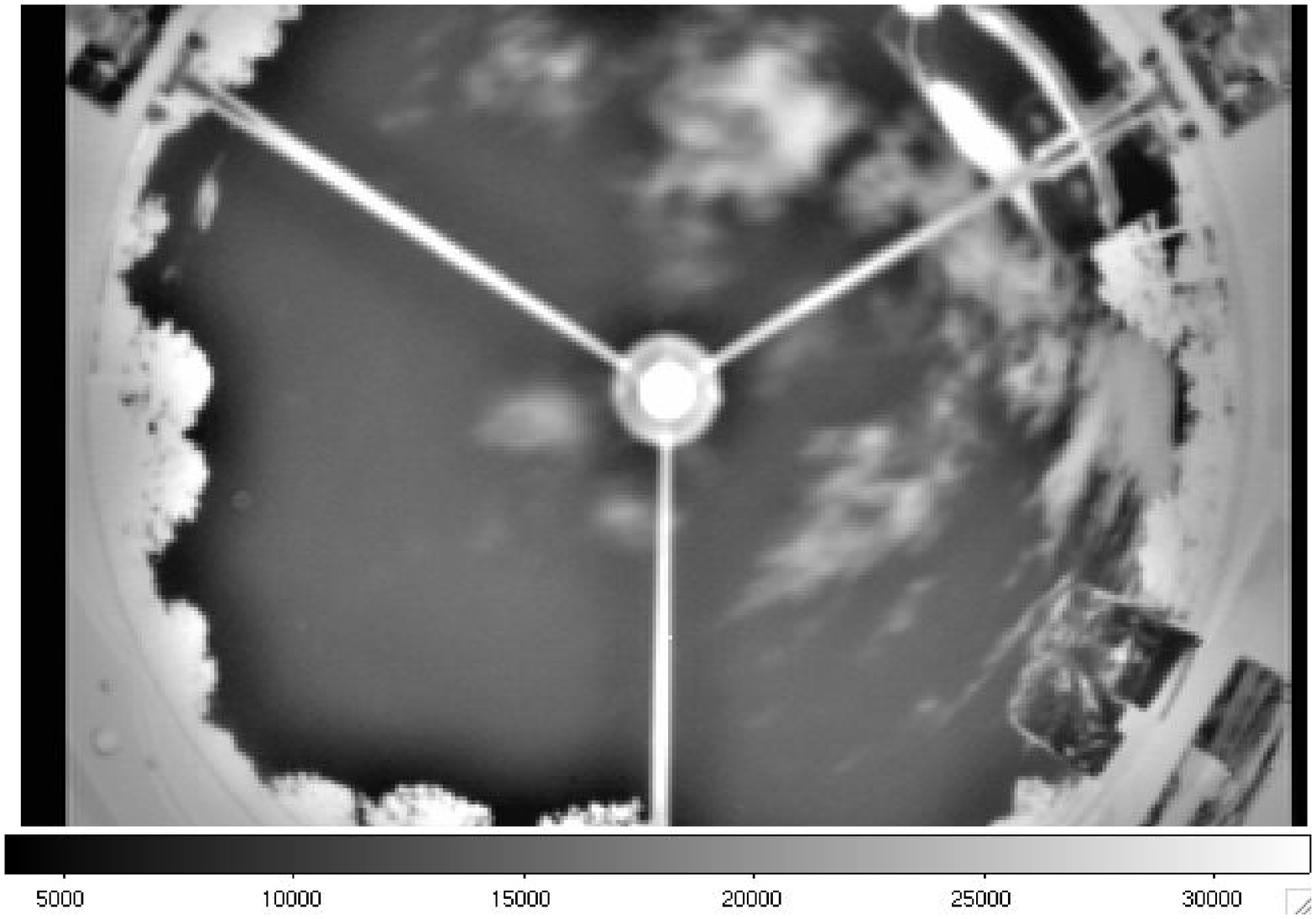}
\caption{Two images from the APO 10 micron all-sky camera:
{\em (left)} clear, {\em (right)} light  clouds.  The bright spidervane
is the camera  support  structure.   (Credit: Astrophysical   Research
Consortium  (ARC)   and the      Sloan Digital  Sky     Survey  (SDSS)
Collaboration, {\tt http://www.sdss.org})
\label{fig5}}
\end{figure}

An excellent resource  for  all-sky  photometric  monitoring is  a  10
micron all-sky camera.   One of  these  has been used successfully  at
Apache Point Observatory (APO), and   has proved essential for  survey
operations and image  quality  assurance for  the SDSS  \citep{Hogg01,
Tucker06}.\footnote{{\tt    http://hoggpt.apo.nmsu.edu/irsc/tonight/}}
The 10 micron  all-sky camera has  also proved useful  for photometric
monitoring for observations on  the Astrophysical  Research Consortium
(ARC) 3.5m telescope, also located at APO.  One of the main advantages
of  the 10 micron  all-sky  camera over an  optical   one is that  the
infrared camera  can detect  even light  cirrus under  a full range of
moon phases  (from  new to   full); see  Fig   $\ref{fig5}$.   Another
infrared  all-sky  camera is  being  built  by Joshua   Bloom and Onsi
Fakhouri     for     the     Pairitel    telescope     at      Whipple
Observatory.\footnote{{\tt
http://astro.berkeley.edu/$\sim$onsi/clic/}} Currently, there is no 10
micron all-sky camera based at CTIO, but there are plans by the DES to
construct one.

Other  resources for monitoring the  photometricity of the sky include
optical   all-sky       cameras          (like     TASCA\footnote{{\tt
http://www.ctio.noao.edu/$\sim$david/tasca.htm}}                   and
CONCam\footnote{{\tt    http://nightskylive.net/cp/}})        and  the
RoboDIMM\footnote{{\tt
http://www.ctio.noao.edu/telescopes/dimm/dimm.html}}   seeing and flux
monitor, which are already available at CTIO.

\section{Nightly Absolute Calibration}

By  nightly absolute calibration, we  mean fitting observed magnitudes
of standard stars to photometric equations for a night's observations.
Our standard star observation strategy  is currently evolving, but  we
believe the general plan will be as follows:
\begin{enumerate}
\item Observe  3  standard star  fields, each  at a  different airmass
      ($X=1-2$), between nautical  (12\deg)  and astronomical (18\deg)
      twilight  (evening  and  morning).    
\item Observe up to 3 more standard star fields (at various airmasses) 
	throughout the night.
\item Also  observe standard star fields  when the  sky is photometric
      but seeing   is too  poor for science   imaging (seeing  $>$ 1.1
      arcsec FWHM).
\item Use fields with multiple standard   stars.
\item Keep an eye on the photometricity monitors.
\end{enumerate}

The plan for  absolute calibration is to  calibrate the photometry  to
the  DES ``natural''   system.   By using the  natural   system of the
telescope$+$camera$+$filters,  no system    response color  terms  are
needed in the photometric  equations, thus simplifying calibration  of
the imaging data as there is no need to couple science images obtained
in different filters.

Due to the similarities in the SDSS and DES filter response functions,
DES calibration can be achieved by using SDSS $u'g'r'i'z'$ and $ugriz$
standards transformed into the  DES $griz$ natural system.  Since  the
SDSS  $g'r'i'z'$/$griz$   and  DES $griz$  filter   responses are very
similar, the transformations from SDSS to DES  should be well behaved.
(Strictly speaking, the  the filter response curves  of the SDSS $z'$,
the SDSS $z$, and the  DES $z$ are not  as similar; so transformations
from  the  SDSS magnitudes   to the  DES magnitudes   may need special
treatment.)

This absolute calibration strategy is similar  to that employed by the
SDSS.  Although  SDSS   photometry  is  published in the    SDSS  2.5m
telescope's $ugriz$ natural  system,  it is in fact   calibrated based
upon  observations of  $u'g'r'i'z'$  standard stars  by the  SDSS 0.5m
Photometric Telescope \citep{Tucker06}.

\section{Standard Stars}

The DES will rely on a set of standard stars to establish its absolute
calibration.   Currently, we plan to   rely on two particular sets  of
standard  stars  that are  now  becoming available  in the SDSS filter
system: the  Southern $u'g'r'i'z'$ standards   and the SDSS  Stripe 82
$ugriz$ standards.

\subsection{The Southern  $u'g'r'i'z'$ Standards} 

One excellent set of standard stars  we plan on  using is the Southern
$u'g'r'i'z'$  standard star  network of \citet{Smith06},  observations
for which were obtained on  the CTIO 0.9m  telescope over the 2000  --
2004 time frame.  This  set of standard stars is  an extension  of the
Northern$+$Equatorial $u'g'r'i'z'$ standard star network that
\citet{Smith02} established for  the  photometric calibration  of  the
SDSS.   The Southern  $u'g'r'i'z'$ standard star   network consists of
approximately  sixty  $13.5\arcmin  \times 13.5\arcmin$  fields.  Each
field  contains typically several   tens   of standard  stars in   the
magnitude  range   $r=9$--18.    Altogether, there    are $\sim$16,000
standards stars  in   Southern $u'g'r'i'z'$  standard  star   network.
(Further  details of the  Southern $u'g'r'i'z'$  standard star network
can be found in Smith's contribution in this volume.)

For the DES, stars as bright as $r \approx  13$ can likely be observed
by DECam with 5$+$ second exposures under conditions of poor seeing or
with de-focusing (FWHM=1.5\arcsec),  making the Southern  $u'g'r'i'z'$
standard  star  network  an  essential  resource for   the photometric
calibration of this survey.

\subsection{SDSS Stripe 82 $ugriz$ Standards}

One  of the priceless  data sets that  the SDSS has provided is Stripe
82, a 2.5\deg-thick region extending along  the celestial equator from
$20^{\rm h}40^{\rm m}$ to $3^{\rm h}20^{\rm m}$ (250 sq deg in total).
Observable   in the  northern  Autumn when    the main  SDSS area (the
Northern Galactic Cap) is inaccessible, Stripe 82 has been imaged $\ga
10\times$  under photometric conditions  by  the  SDSS 2.5m  telescope
\citep{Adelman-McCarthy07}.  The quality and multi-epoch nature of the
Stripe 82 data make  them an ideal source  for creating  tertiary SDSS
$ugriz$ standard stars.   An effort by  Ivezi\'c and collaborators has
already   yielded a Stripe   82    standard stars  catalog  containing
$\sim10^6$  $ugriz$  tertiary  standards   ---  on  average $\sim$4000
standards per  sq deg ---   in the magnitude range   of $r =$ 14.5--21
(\citealt{Ivezic06}; also see Ivezi\'c's contribution in this volume).

We  note that the SDSS  Stripe 82 is already  a part of the DES survey
region (recall  Fig.~\ref{fig2}). It is readily  observable at a range
of airmasses  throughout most nights during  the DES program,  and the
2.5\deg  width    of  Stripe  82  compares     favorably  with DECam's
2.2\deg-diameter  field-of-view.  A Stripe   82 standard stars catalog
will therefore play a critical  role in the photometric calibration of
the DES.

\subsection{Other Sources of Standard Stars}

As the SDSS  $ugriz$ filters becomes  more popular, other large  scale
projects are   making use of  them  and even  are  creating  their own
standard star catalogs for these filters.  These include standards for
the VST OmegaCam and  for the SkyMapper  survey (see contributions  in
this volume by Kleign and Bessell, respectively).  These other sources
of standards may also prove useful for the DES.

\section{Global Relative Calibrations}

Global Relative Calibrations cover    two topics: the removal  of  any
residual photometric effects due  to vignetting and/or scattered light
(star flat corrections),     and the removal of  any    field-to-field
(``Hex-to-Hex'')    photometric  zeropoint offsets   due  to observing
different fields  under   different photometric conditions.   Let   us
consider each in turn.

\subsection{Star Flats}

Due to   vignetting and stray  light, a  detector's  response function
differs for point sources and  extended sources. Standard flat  fields
(domes, twilights, skies)  may flatten an  image sky  background well,
but not necessarily the stellar photometry.  This is particularly true
of wide-field imagers,  in which these  effects can produce systematic
photometric   errors   of   10--20\%   or  more   across  an  imager's
field-of-view.   The solution is    to create  and apply  star   flats
\citep{Manfroid95}.   A star flat can be  created simply by offsetting
an uncalibrated field (like an  open cluster) multiple times and  then
fitting a spatial function  to  the magnitude differences  for matched
stars  from the different  exposures  \citep{Manfroid01}.  Likewise, a
star flat can be created by observing a well-calibrated field once and
then fitting a spatial  function to the observed$-$standard magnitudes
for the stars in the field \citep{Manfroid96,  Koch04}; this of course
assumes the availability  of a densely  populated standard  star field
that is at least  as wide as   the imager's field-of-view.  In  either
case, however  it  is  created, the  star flat   can then be   applied
directly to the measured photometry.

The  DES can profit  from both methods  of star flat creation.  First,
due to its  survey strategy of  large offsets between each tiling (see
Section 9.2  below), star flat  correction maps can  be generated from
previously all the uncalibrated stars in the survey area at the end of
the survey using the information from the multiple tilings of the full
survey region.  Second, observations of SDSS Stripe  82 (and its dense
population   of  tertiary $ugriz$  standard   stars) during normal DES
operations  will permit the creation of  high quality  star flats very
early within  the  survey.  Since fields   within SDSS  Stripe  82 are
targetted throughout the 5-year course of DES operations, we will also
be able to track any evolution of the star flats themselves over time.

\subsection{Hex-to-Hex Zeropoint Offsets}

\begin{figure}    
\centering
\vspace{-.25cm}
\plotone{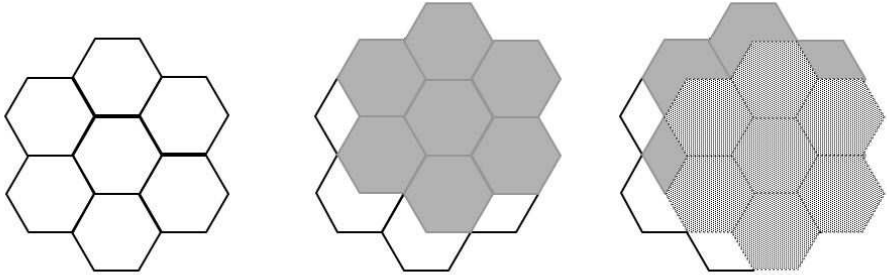}
\caption{Hex offset tiling strategy:  
{\em (left)} 1 tiling, {\em (middle)} 2 tilings, {\em (right)} 3 tilings. 
(Credit:  Jim Annis)
\label{fig6}}
\end{figure}

Recall from Section 3 that the current  idea for survey strategy is to
cover the entire 5000 sq  deg of the survey  area  twice per year  per
filter.  Each full coverage of the sky in a filter is called a tiling.
The  goal is, after 5  years, to have tiled  the survey region 5 times
each in $g$ and $r$, 7 times in $i$, and 13 times  in $z$.  For a 3 sq
deg  wide-field  camera, it takes $\sim$1700  hexes  to tile the whole
survey area.

Our   recipe  for  covering the  survey  region   is  as  follows (see
Fig.~\ref{fig6}):
\begin{enumerate}
\item Tile the survey region.
\item Then, tile the survey region again  with the hexes offset half hex over and up
	(this gives 30\% overlap with three hexes).
\item Repeat, with different offsets.
\end{enumerate}
We  note  that  this recipe  is   similar to  PanStarrs  strategy (see
Magnier's contribution in   this volume).  The large  overlaps provide
very   robust hex-to-hex   relative  calibrations.   Relative  offsets
between tiles can be solved for using a large matrix inversion.

\subsection{Simulation}

\begin{figure}    
\centering
\vspace{-.25cm}
\plotone{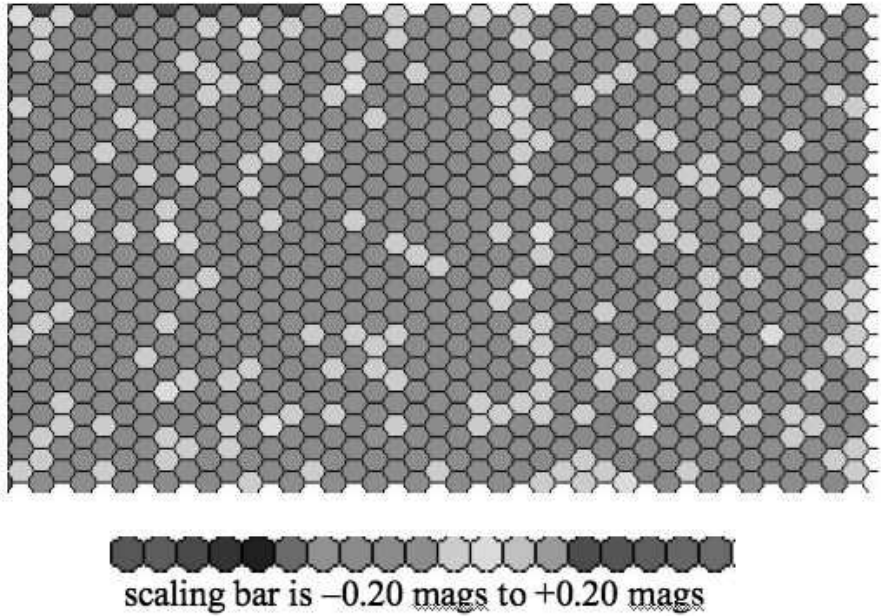}
\caption{Simulated map of the photometric calibration error using relative 
photometry.  The  full range  of  the scaling  bar  is $-$0.20  mag to
$+$0.20 mag. the map is that resulting after 3 tilings  and has an rms
scatter of $\sigma=0.013$  mag,  starting from  10\% photometry and  a
variety of  flatfielding and scattered   light systematics embedded at
the $\sigma=0.03$ mag level.(Credit: Jim Annis \& Huan Lin)
\label{fig7}}
\end{figure}

To  test whether our recipe of  multiple offset  tilings could achieve
our  survey requirements  of  2\%  (0.02 mag)  all-sky photometry,  we
performed a simple survey simulation.   For sky conditions, we assumed
$\approx$10\% variations in photometric zeropoints for the hexes.  For
the camera itself, we included a multiplicative flat field gradient of
amplitude 3\% from  east to west across the  focal plane, an  additive
scattered light pattern with a $1/r^2$ amplitude from the optical axis
(reaching 3\% at the   edge of the focal   plane), and  an  additional
additive  3\% rms scattered light    per individual CCD  on the  focal
plane.   We find that  we can  indeed  achieve our survey requirements
after only  2 tilings; in fact, we  can even achieve our enhanced goal
of 1\%  (0.01 mag) all-sky photometry   after 5 complete  tilings (see
Table~1 and Fig.~\ref{fig7}).

\begin{table}
\caption{Simulation Results for Relative Calibration}
\smallskip
\begin{center}
{\small 
\begin{tabular}{c c}
\tableline
\noalign{\smallskip}
Tiling & $\sigma$\\
       & [mag] \\ 
\noalign{\smallskip}
\tableline
\noalign{\smallskip}
1 & 0.035 \\
2 &  0.18 \\
5 & 0.010 \\
\noalign{\smallskip}
\tableline
\end{tabular}
}
\end{center}
\end{table}

\section{Global Absolute Calibration}

For  a  firm Global  Absolute   Calibration,  we  need  one   or  more
spectrophotometric standard stars which have been calibrated (directly
or indirectly) to a  NIST standard source  and an accurately  measured
total system  response for each filter  passband for at least one CCD,
including filter  transmissions, CCD QE responses, optical throughput,
and  atmospheric transmission.  We  then perform  synthetic photometry
for   each   of  these   spectrophotometric   standard  stars,  simply
calculating  the expected  photon flux $F_{\rm   exp}$  in each filter
passband.  Then, we  measure the magnitude for each spectrophotometric
standard in each  filter passband with  the Blanco$+$DECam.   Finally,
for each passband, we can compare the expected magnitude $m_{\rm exp}$
for each standard based  upon its expected  photon flux $F_{\rm  exp}$
against its observed magnitude $m_{\rm  obs}$; the resulting zeropoint
offset can then be applied to the photometry for absolute calibration.

The DES filter transmissions, the CCD QEs,  and the optical throughput
for the Blanco$+$DECam can be  measured via a monochromator.  (Another
possibility to  measure these is via  a tunable dye laser flatfielding
system;  please see  Chris  Stubbs', David Burke's,  and Yorke Brown's
contributions to this volume.)   The atmospheric transmission spectrum
for      CTIO         has        already           been       measured
\citep{Stone83,Baldwin84,Hamuy92,Hamuy94}.     Furthermore,    several
potentially useful  spectrophotometric standards are  available (e.g.,
GD~71,   G158-100,   GD~50,  and  G162-66).   All     are white  dwarf
spectrophotometric standards, all are  visible from CTIO, and  all are
faint enought  ($V \ge 13.0$)that they will  not saturate the DECam at
exposure times   of 5  sec  for  seeing/focus   of  1.5 arcsec   FWHM.
Therefore, most of the information  for Global Absolute Calibration is
either already  in hand (e.g.,   CTIO atmospheric transmission spectra
and well-calibrated spectrophotometric  standards) or  will eventually
be available (e.g.,  DES filter transmission curve measurements, DECam
CCD QEs, Blanco$+$DECam optical throughputs.)

\section{Conclusions}

We have discussed  the Dark Energy  Survey (DES) and the current plans
for its photometric calibration.  To    achieve the survey's   science
goals, all-sky photometry  of 2\% (0.02  mag) rms is required, with an
enhanced  goal  of   1\%  (0.01) rms.    This   is  a challenging  but
achieveable goal.

\acknowledgments
The FNAL contribution to this work was supported by the US Department of 
Energy under contract DE-AC02-76CH03000.

\end{document}